\documentclass[a4paper]{jpconf}
\usepackage{graphicx}
\usepackage{citesort}
\begin{document}

\bibliographystyle{iopart-num}

\def\registered{{\ooalign{\hfil\raise .00ex\hbox{\scriptsize R}\hfil\crcr\mathhexbox20D}}}
\def\numberzero{{\ooalign{\hfil\raise .00ex\hbox{\scriptsize 0}\hfil\crcr\mathhexbox20D}}}
\def\numberone{{\ooalign{\hfil\raise .00ex\hbox{\scriptsize 1}\hfil\crcr\mathhexbox20D}}}
\def\numbertwo{{\ooalign{\hfil\raise .00ex\hbox{\scriptsize 2}\hfil\crcr\mathhexbox20D}}}
\def\numberthree{{\ooalign{\hfil\raise .00ex\hbox{\scriptsize 3}\hfil\crcr\mathhexbox20D}}}
\def\numberfour{{\ooalign{\hfil\raise .00ex\hbox{\scriptsize 4}\hfil\crcr\mathhexbox20D}}}
\def\numberfive{{\ooalign{\hfil\raise .00ex\hbox{\scriptsize 5}\hfil\crcr\mathhexbox20D}}}
\def\numbersix{{\ooalign{\hfil\raise .00ex\hbox{\scriptsize 6}\hfil\crcr\mathhexbox20D}}}
\def\numberseven{{\ooalign{\hfil\raise .00ex\hbox{\scriptsize 7}\hfil\crcr\mathhexbox20D}}}

\title{The upgrade of GEO\,600}

\author{ H.~L\"uck$^1$, C.~Affeldt$^1$, J.~Degallaix$^1$, A.~Freise$^2$, H.~Grote$^1$, M.~Hewitson$^1$, S.~Hild$^{3}$, J.~Leong$^1$, M.~Prijatelj$^1$, K.A.~Strain$^{1,3}$, B.~Willke$^1$, H.~Wittel$^1$ and K.~Danzmann$^1$}

\address{$^1$ Max-Planck-Institut f\"{u}r Gravitationsphysik (Albert-Einstein-Institut) and Leibniz Universit\"{a}t Hannover,
Callinstr. 38, 30167 Hannover, Germany\\
$^2$ School of Physics and Astronomy, The University of Birmingham, Edgbaston, Birmingham, B15 2TT, United Kingdom\\
$^3$ Institute for Gravitational Research, University of Glasgow, Glasgow, G12 8QQ, United Kingdom, United Kingdom}

\address{Institut f\"ur Gravitationsphysik, Leibniz Universit\"at Hannover, Callinstr. 38, 30167 Hannover, Germany}

\ead{harald.lueck@aei.mpg.de}

\begin{abstract}
The German\,/\,British gravitational wave detector GEO\,600 is in the process of being upgraded. The upgrading process of GEO\,600, called GEO-HF, will concentrate on the improvement of the sensitivity for high frequency signals and the demonstration of advanced technologies. In the years 2009 to 2011 the detector will undergo a series of upgrade steps, which are described in this paper.
\end{abstract}

\section{Introduction}
GEO\,600, the British\,/\,German gravitational wave (GW) detector\cite{GEO600}, is one of the worldwide network of GW detectors. The main initial goal of GW detectors is the first detection of GW signals. In order to improve the chances for detecting a signal, the world-wide collaboration of GW scientists typically operates more than one detector in coincidence. GEO\,600 finished its latest data run in July 2009 after taking data for about 3.5 years, partly together with LIGO\cite{LIGO} and Virgo\cite{Virgo}. To further improve the chances of detecting a signal the detectors have been - and will further be - upgraded to increase their sensitivity. The first upgrades for the LIGO and Virgo detectors to the 'enhanced' versions happened from late 2007 to July 2009. In this period, called Astrowatch, GEO\,600 took data to cover the downtime of the two 4\,km LIGO detectors and the Virgo detector. In the Astrowatch period GEO\,600 was partly joined in data taking by the 2\,km LIGO detector located near Hanford, Washington, forming a small detector network. The performance of GEO\,600, and GEO\,600 related activities during Astrowatch are described elsewhere\cite{HartmutGEO600Amaldi}. Rather than just detecting that there was a signal at all, the goal of future generations of GW detectors is to do GW astronomy, i.e. to focus on extracting astro-physically interesting science from the GW signals. The upgrades of GEO\,600, which are described in this paper, will firstly enable GEO\,600 to take data during a part of the upgrading time of LIGO and Virgo to the 'advanced' state with scientifically interesting sensitivity and secondly to test and prove advanced technologies that can be used in future generations of GW detectors, e.g. Squeezing. This paper outlines the possibilities, and plans to enhance the sensitivity of the GEO detector. Section \ref{status} gives a short overview of the GEO\,600 configuration before the upgrades. The main steps in the upgrade will be the injection of squeezed states into the output port, tuned Signal Recycling with DC readout, installation of an output mode cleaner, the change of the Signal Recycling bandwidth and a power increase inside the interferometer in combination with the installation of a thermal compensation system. The individual steps of the upgrading process are described in section \ref{upgrades}. The sensitivity improvement mainly concerns the {\bf h}igh {\bf f}requency part, which is why the upgrading process is called GEO-{\bf HF}. At lower frequencies thermal noises become a limit which cannot significantly be lowered with currently available techniques.

\section{Status Quo}\label{status}
GEO\,600 is a dual recycled (i.e. Power- and Signal-Recycled) Michelson type interferometer with an arm length of 600\,m. The arms are folded once giving an effective arm length of 1200\,m.

\begin{figure}[h]
\centering\includegraphics[width=\textwidth]{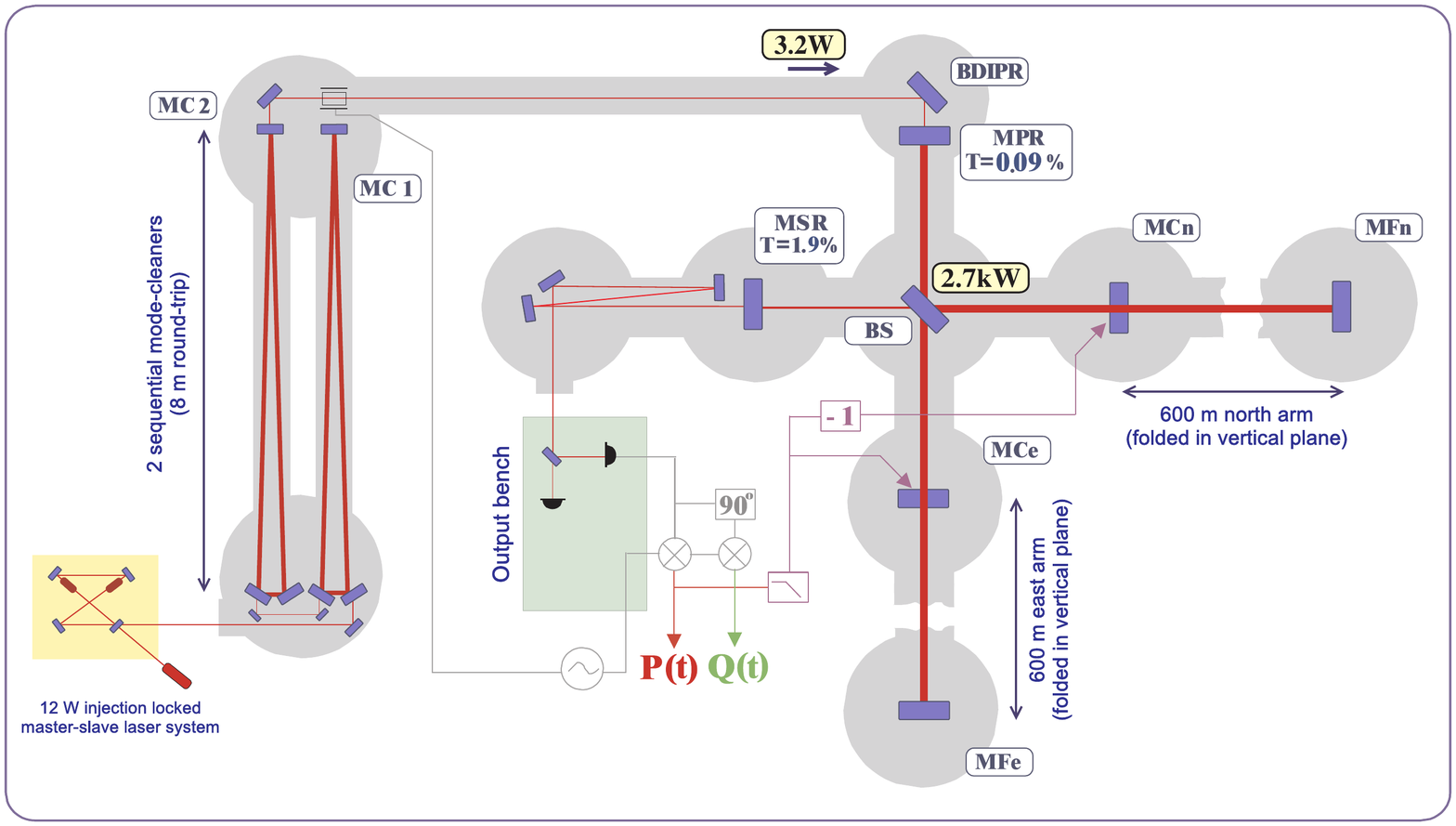}
\caption{\label{layout}Schematic layout of GEO\,600}
\label{GEOlayout}
\end{figure}

\begin{figure}[h]
\centering\includegraphics[width=0.7\textwidth]{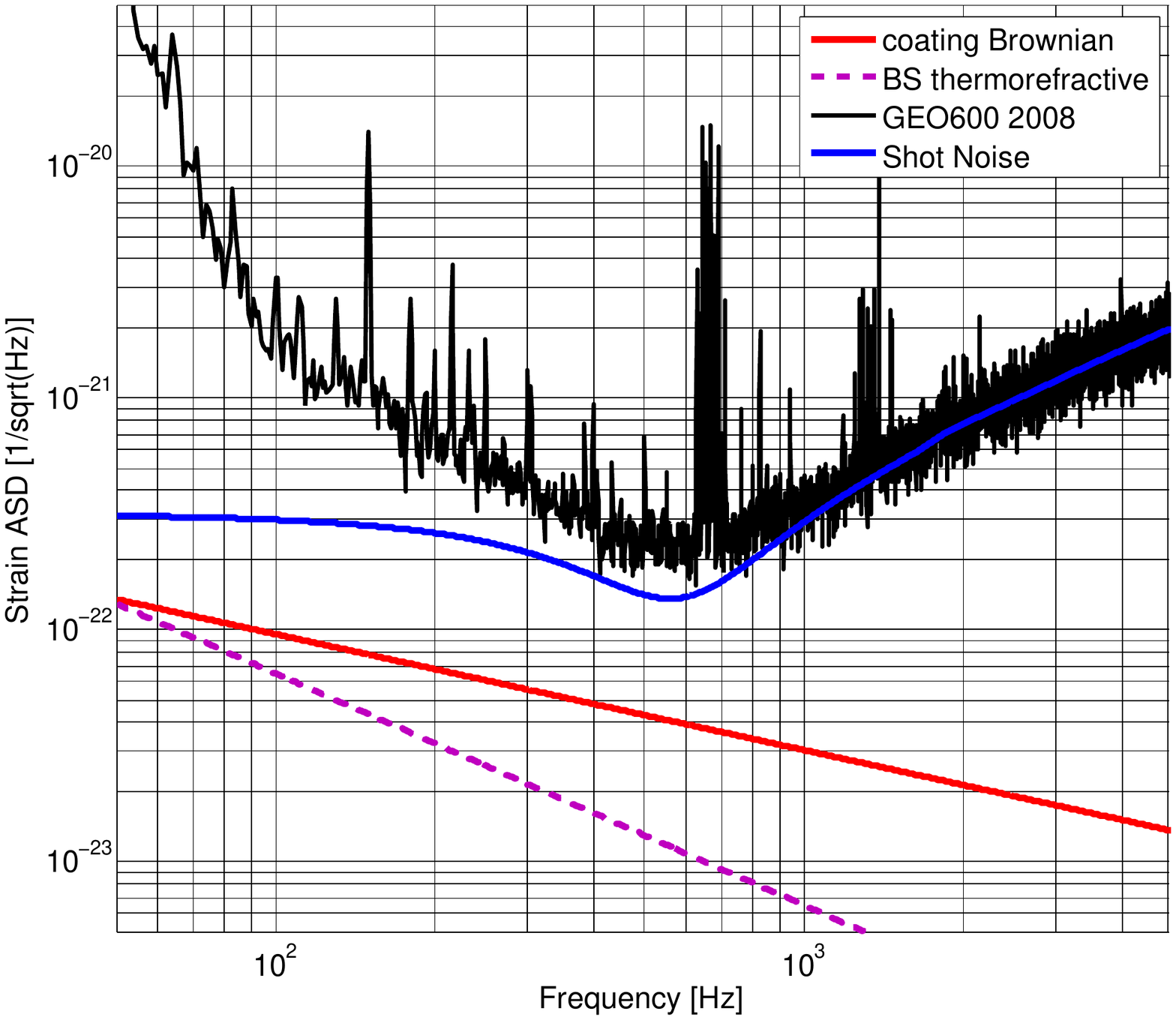}
\caption{Sensitivity of GEO\,600}
\label{GEOSensitivity}
\end{figure}

In the past few years GEO\,600 used Schnupp modulation\cite{Schnupp}, see also section\ref{RF2DC}, for a heterodyne read--out, Signal Recycling detuned to a frequency of 530\,Hz, a light power at the input to the Power Recycling cavity of about 3\,W with a Power Recycling factor of about 1000. A schematic layout with the relevant parameters is given in figure\,\ref{GEOlayout}. The light from a 12\,W Nd-YAG laser with a wavelength of 1064\,nm is filtered by two optical resonators, the Mode Cleaners, and then sent through modulators and Faraday isolators to the Power Recycling cavity. The peak sensitivity of GEO\,600, $2.2\cdot10^{-22} [1/\sqrt{Hz}]$, is reached at the tuning frequency of 530\,Hz (see figure\,\ref{GEOSensitivity}). Above this frequency the sensitivity of GEO\,600 is mostly shot noise limited. Between roughly 100\,Hz and 500\,Hz a yet unknown noise source limits the sensitivity in 'detuned RF readout', whereas below 100\,Hz technical alignment related noise sources dominate the spectrum. The latest status of recent GEO commissioning is described in \cite{HartmutGEO600Amaldi}. In recent 'tuned DC readout' experiments all of the observed noise could be explained by simulated (e.g. thermal noise, quantum noise) and measured (e.g. alignment feedback noise) contributions from known sources. 

\section{Upgrading plan}\label{upgrades}
The upgrade of GEO\,600 will be done in a staged way, checking the performance after each individual step. The upgrading is aimed at frequencies above 500\,Hz where roughly an order of magnitude can be gained in sensitivity by lowering shot noise. Below 500\,Hz GEO\,600 will be limited by thermal noises, mainly thermo-refractive noise at the beam splitter and coating Brownian noise. Techniques to reduce these noise sources are available\cite{LaguerreGauss},\cite{Nonquarterstacks},\cite{Titania} but are beyond the scope of the envisioned GEO upgrading plans for the next few years. The upgrade will therefore concentrate on lowering the shot noise.\par
Shot noise in an interferometric gravitational wave detector originates mainly from the vacuum fluctuations entering the output port. The contributions that come from the fluctuations on the input laser beam are small compared to those entering the output port because the input laser beam is well stabilized, filtered by the Power Recycling cavity and the Schnupp asymmetry and the imbalance in the losses of the arms of the interferometer are small. Reducing the vacuum fluctuations entering the output port in the phase quadrature can therefore reduce the observed shot noise, see section \ref{Squeezing}.

\subsection{From RF readout to DC readout}
\label{RF2DC}
The GEO\,600 interferometer is held at a 'dark fringe' operating point. At this operation point the arm length difference is arranged such that the light fields returning from the interferometer arms interfere destructively at the output port; hence it is {\bf dark}. A control system maintains this condition. Right on the 'dark fringe' there is no dependence of the light power at the output port on the arm length difference, hence the light power can neither be used as an error signal for the control system nor as a signal sensing GWs. Differentially modulating the arm length around the interference minimum and demodulating the resulting power detected at the output port solves this problem. A physical modulation of the mirror position would introduce too much noise. So called 'Schnupp modulation'  opens a way around this problem. If the laser light is frequency modulated and if the arm lengths are slightly different (13.5\,cm in the GEO\,600 case) the same effect as differentially modulating the arms lengths is achieved. In GEO\,600 the modulation frequency is about 14.9\,MHz and therefore the readout technique is called {\bf R}(adio)-{\bf F}(requency) readout or heterodyne readout. Unfortunately this method collects additional vacuum noise from twice the modulation frequency, which, beating with the modulation sidebands at the modulation frequency, produces noise at the modulation frequency which gets demodulated into the signal frequency band\cite{Buonanno}. This RF-readout technique was chosen for the first generation of GW detectors because the lasers used were shot noise limited only in the MHz frequency range. The excellent amplitude and frequency stability of (stabilized) solid state ND-YAG lasers in the detection frequency band in addition to the filtering properties of the Power Recycling cavity now allows to switch to a self--homodyne readout technique, also called DC readout. In this case the operation point is slightly (in the range of about 5\,pm to 50\,pm) detuned from the dark fringe, which yields a dependence of the light power at the output port on the arm length difference. The shot noise for DC readout is lower than for RF readout. All large interferometric gravitational-wave detectors will or have switched to DC readout with the upgrade from the first generation to the next upgrade step. The first experiences with DC readout in the GEO\,600 detector are reported in \cite{HartmutGEO600Amaldi},\cite{DCreadoutGEO} and \cite{JeromeAmaldi}. DC readout is also beneficial for the injection of squeezed states as only the correct phasing at the laser carrier frequency and the respective signal sidebands has to be taken into account; see also section\,\ref{Squeezing}.\par
In GEO\,600 Schnupp modulation will still be used for generating the error signals of the differential wave--front sensing for the Auto--Alignment system. As a status of January 2010 GEO\,600 is routinely operating with DC readout. 

\subsection{\label{OMC} Output Mode Cleaner}

\begin{figure}[h]
\centering\includegraphics[width=0.7\textwidth, bb=0 0 572 384]{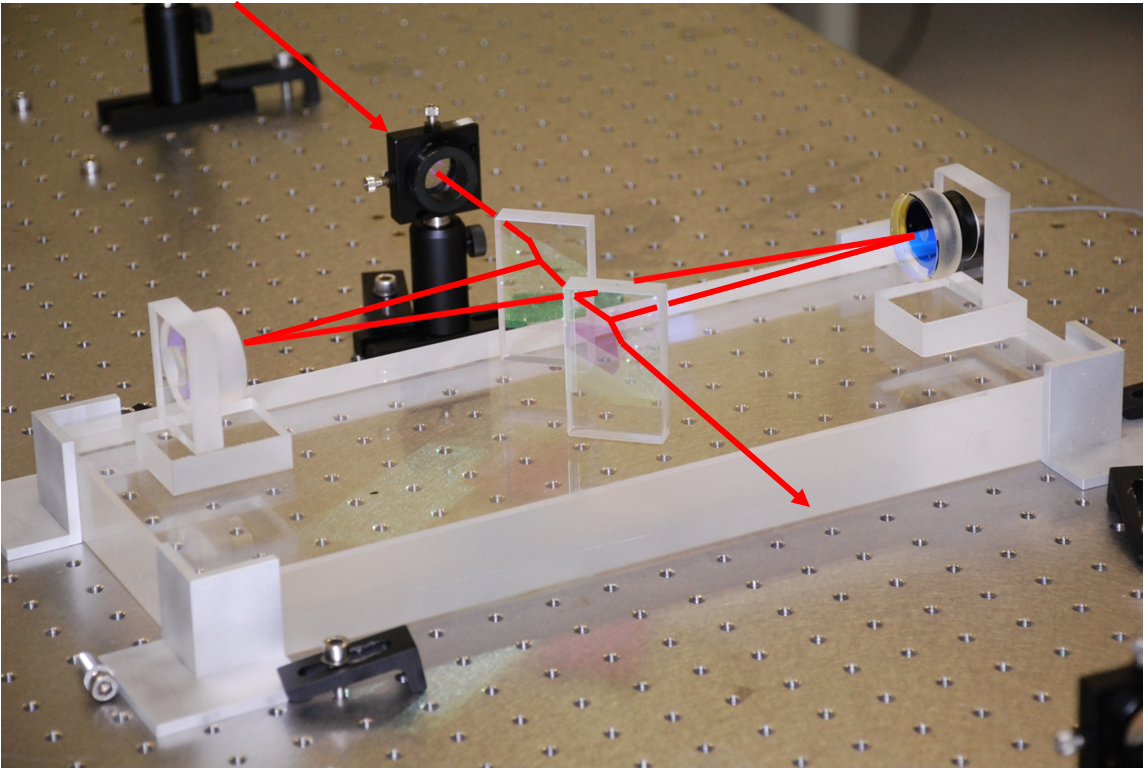}
\caption{\label{OMCfigure}The output mode cleaner for GEO\,600}
\end{figure}

The light leaving the output port of GEO\,600 will be sent through an Output Mode Cleaner (OMC)\cite{JeromeAmaldi} before being detected on the photo-detector. The OMC is a four mirror cavity with a round trip length of roughly 66\,cm as shown in figure\,\ref{OMCfigure}. The mirrors are glued onto a fused silica base plate to reduced thermal expansion to a level which allows the use of a small range ($\approx3 \,\mu m$) piezo-electric actuator. The Finesse of about 150, giving a FWHM of about 3\,MHz, assures that the modulation sidebands (14.9\,MHz) are reflected by the OMC when it is tuned to carrier resonance. With proper mode matching the light in the TEM--00 mode will get transmitted and the unwanted light in higher TEM modes will be reflected. One of the mirrors is mounted on a piezo-electric actuator. Modulating the OMC length with this Piezo-actuator allows the generation of error signals for the length control and for the Auto--alignment system aligning the beam onto the OMC. More details on this technique can be found in \cite{OMCMirko}. At the end of January 2010 the OMC is installed and stably operating. 

\subsection{\label{Squeezing} Squeezing}
With the shot noise contribution being the main noise source limiting the GEO600 sensitivity, the upgrade plans are aiming at lowering the relative shot noise level. The interferometer is kept close to a 'dark fringe' operating point (DC readout; see section \ref{RF2DC}) where the output port is almost dark. For energy conservation reasons all the light must be reflected towards the input port. For symmetry reasons the same must be valid for fields entering the output port. The vacuum fluctuations entering the output port therefore get reflected from the interferometer and interfere with the carrier light getting to the output port due to the offset from the dark fringe for DC readout. As it is the fluctuations in the phase quadrature that cause the shot noise, the injection of squeezed light, where the noise in the phase quadrature is decreased, can lower the shot noise detected by the photo detector. GEO\,600 will use a squeezed vacuum field with a squeezing level of about 10dB to be injected into the output port. The injection will be done via a Faraday isolator in an additional vacuum vessel added to the GEO vacuum system. This vessel contains some mode matching optics for the Output Mode Cleaner, the Output Mode Cleaner itself, a Faraday isolator and a photo diode. The vacuum level is about $5\cdot10^{-3}$\,mbar. All the optical elements are placed on a platform seismically isolated with a Minus-K$^\registered$ isolation system.\par
\begin{figure}[h]
\centering\includegraphics[width=\textwidth]{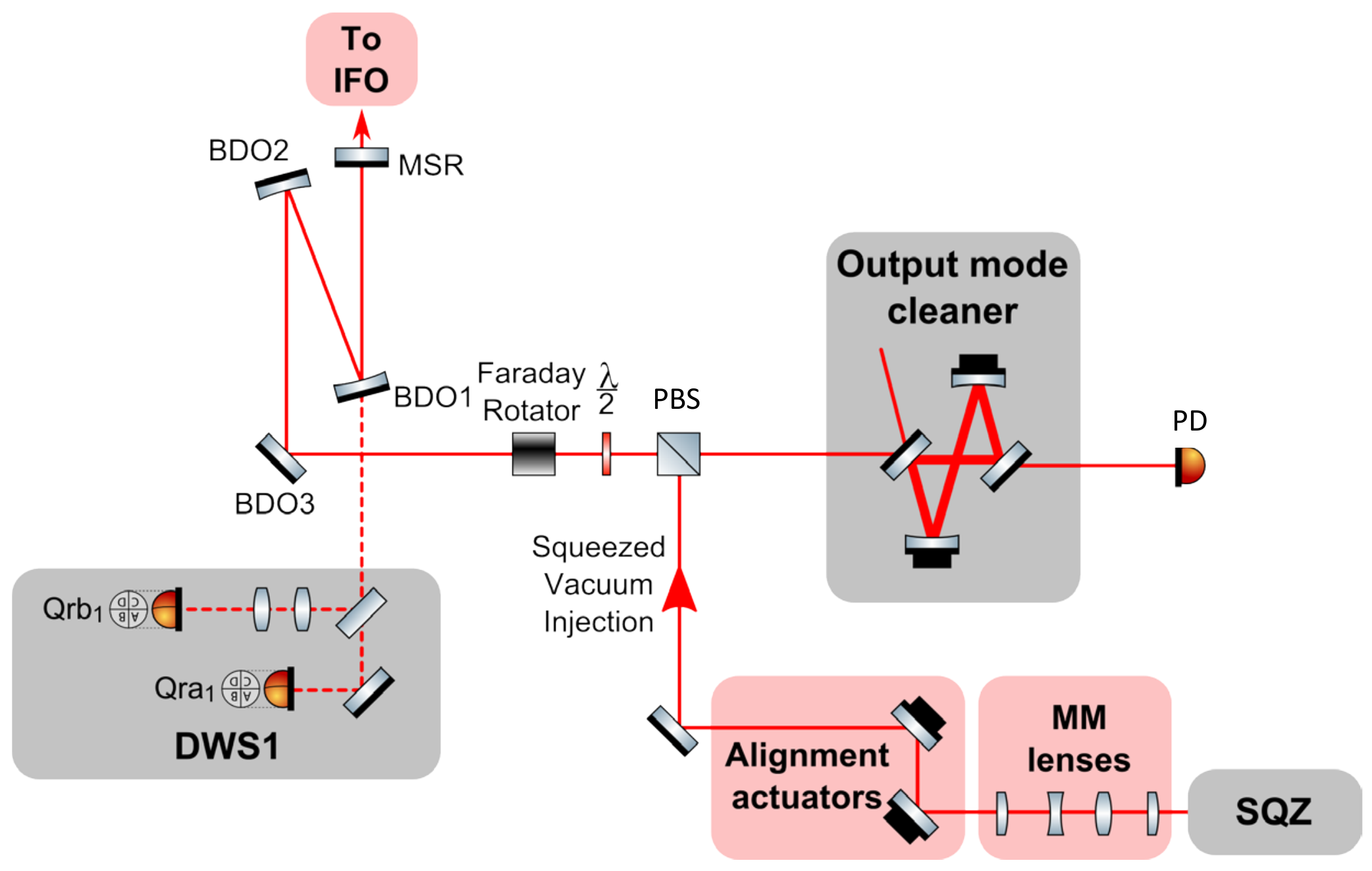}
\caption{\label{squeezingfig}Schematic layout of the output optics of GEO\,600, showing the squeezing input and the output mode cleaner. The output signal of the interferometer is transmitted through the Signal Recycling mirror (MSR) and de-magnified by the curved mirror BDO1. BDO2 and BDO3 serve as steering mirrors aligning the output mode to the Eigenmode of the OMC. The half-wave plate, the Faraday rotator and the polarizing beam splitter (PBS) serve as an optical diode to inject the squeezing from the one direction and let the output signal pass through the OMC to the main photo diode (PD) in the other direction. The mode matching lenses and the beam steerers between the PBS and the squeezing system align and match the squeezing mode to the interferometer mode. The differential wave-front sensing system (DWS1) gives the error signals for aligning the two beams of the interferometer arms to each other.}
\end{figure}

To make optimal use of the squeezing over the full detection frequency range, the squeezing ellipse, i.e. the phase of the injected squeezed light, must be optimally oriented with respect to the light leaving the output port for all frequencies within the detection bandwidth. In case of a detuned (with respect to the carrier) Signal Recycling cavity this can be achieved by sending the squeezed light through additional filter cavities\cite{FilteredSqueezing} or using Twin Signal Recycling\cite{TwinSigRec}. For tuned (zero frequency detuning) Signal Recycling the orientation of the squeezing ellipse is frequency independent and no filtering of the squeezed light is required. The improvement in sensitivity by using squeezing (together with the series of other changes) is shown in figure \ref{Sensitivities}. In January 2010 the squeezing breadboard, which provides the squeezed input for GEO\,600, is providing strong, broadband squeezing, is undergoing some long term stability tests and will be brought to the site in late February. The additional vacuum vessel with all its components is in place and operative. A schematic layout of the output section of GEO\,600 is shown in figure\ref{squeezingfig}

\subsection{\label{SignalRecycling}Tuned Signal Recycling}
\begin{figure}[h]
\centering\includegraphics[width=\textwidth, bb=0 0 800 600]{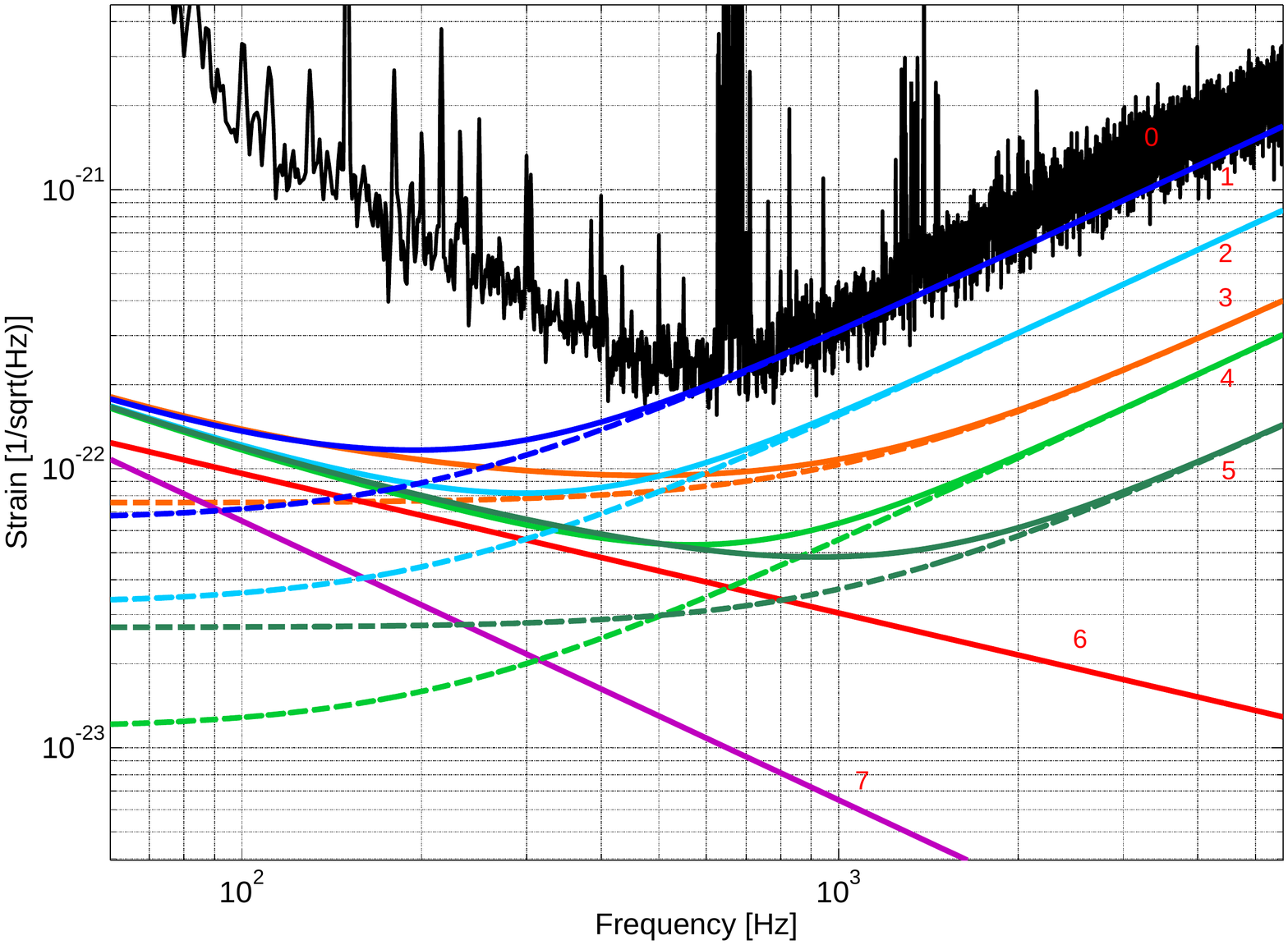}
\caption{\label{Sensitivities}Sensitivity of GEO\,600 for different  configurations.\\
Solid lines indicate total noise, dashed lines represent shot noise only. Technical noise sources are neglected here.\\
\numberzero\,GEO\,600 sensitivity 2009\\ 
\numberone\,DC, tuned SR, $T_{SR}$=2\%, P=3.2\,W\\
\numbertwo \,DC, tuned SR, $T_{SR}$=2\%, P=3.2\,W, 6\,dB Squeezing\\
\numberthree \,DC, tuned SR, $T_{SR}$=10\%, P=3.2\,W, 6\,dB Squeezing\\
\numberfour \,DC, tuned SR, $T_{SR}$=2\%, P=20\,W, 6\,dB Squeezing\\
\numberfive \,DC, tuned SR, $T_{SR}$=10\%, P=20\,W, 6\,dB Squeezing\\
\numbersix \,Coating thermal noise\\
\numberseven \,Thermo-refractive noise of the beam splitter}
\end{figure}

In the future GEO\,600 will use tuned Signal Recycling together with the squeezing to improve the sensitivity over the full frequency range\cite{TunedSR}. In this case the Signal Recycling mirror is shifted such that the Signal Recycling cavity is resonant for the laser carrier light instead of the 530\,Hz signal sidebands as in the detuned mode GEO\,600 operated in before. The achievable sensitivity is shown in figure \ref{Sensitivities}. Tuned Signal Recycling resonantly enhances both signal sidebands, providing higher optical gain on resonance, which results in lower shot noise at the resonance frequency than detuned Signal Recycling. In the case of GEO\,600 this does not influence the total sensitivity as at low frequencies the total noise is dominated by thermal noise and technical noises anyway. 

\subsection{\label{SRTransmission}Increasing the Signal Recycling Transmission}
Widening the bandwidth of GEO\,600 by increasing the transmission of the Signal Recycling mirror from 2\% to 10\% improves the high frequency shot noise at the expense of the low frequency one. Being limited by thermal noises at frequencies below 700 Hz anyway this influences the overall noise for low frquencies only marginally. The crossover point between coating thermal noise and shot noise only moves from 700\,Hz (curve\,\numberfour\,in figure\,\ref{Sensitivities}) to 800\,Hz (curve\,\numberfive\,in figure\,\ref{Sensitivities}), while the high frequency total noise drops by a factor of about two. A further increase in bandwidth would raise the low frequency shot noise to an undesirable level while giving marginal improvement at high frequencies.

\subsection{\label{PowerUp}Power Increase}
The final step in the series of GEO\,600 upgrades, foreseen to be done in late 2010 and in 2011, is an increase in light power inside the interferometer by a factor of about ten. To achieve this several changes become necessary. In the following two sections the changes that will lead to a power increase in the interferometer by a factor of about 8 will be described. The modifications shown in sections \ref{LocalControls} and \ref{TCS} are needed to stably operate the detector at a power level of about 20\,kW in the Power Recycling cavity. 

\subsubsection{\label{Laser}Laser Power increase}
Currently GEO\,600 is operated using a 10\,W master-slave Nd-YAG laser with a wavelength of 1064\,nm. In the current operation mode the laser is attenuated to about 6\,W before being sent into the vacuum system. Due to losses in the Mode Cleaners (see below) only about 3.2W arrive at the Power Recycling mirror as indicated in figure\ref{layout}. The laser will be exchanged for a master-slave-amplifier system delivering about 30\,W, which has been developed by the Laser Zentrum Hannover. It is the same kind of laser that is being used in the 'enhanced' versions of the other large gravitational wave detectors Enhanced LIGO\cite{EnhLIGO} and Virgo+\cite{Virgo+}. The increase in laser power will yield a power increase of about 5 in the Power Recycling cavity.

\subsubsection{\label{MCMod}Mode Cleaner Modifications}
Due to scattering losses of the mirrors of the GEO Mode Cleaners (i.e. two optical resonators, which remove higher order TEM modes from the laser beam and reduce angular beam fluctuations) about 50\% of the light is lost in the Mode Cleaners. At the current level of laser power, radiation pressure already needs to be compensated when bringing the GEO Mode Cleaners into resonance. Hence increasing the laser power would lead to undesirable problems with radiation pressure in the Mode Cleaners when keeping the same cavity finesse. Reducing the reflectivity of the coupling mirrors and with it the power build-up inside the Mode Cleaners, improves the throughput and reduces radiation pressure problems. The filtering properties of the initital Mode Cleaners were very conservative\cite{coreoptics} and will still yield a sufficiently high suppression level after the decrease in finesse\cite{coreoptics}. This mirror exchange will give a power increase in the Power Recycling cavity of a little less than two. The high-power compatibility of the auxiliary optics still needs to be confirmed. 

\subsubsection{\label{LocalControls}Local Control Changes}
The mechanical resonances of the triple pendula for seismic isolation of the main interferometer mirrors are actively damped by a local control system using shadow sensors to monitor the movement of the uppermost mass\cite{GEOsusp}. These shadow sensors are operated with a DC light source and unfortunately detect the light scattered by the interferometer mirrors. Hence slight fluctuations in the circulating light power, which can originate from a slight misalignment of one of the interferometer mirrors, results in a signal at the shadow sensor which is regarded as a movement of the upper mass and a 'compensation' signal is fed back to the actuator. This misaligns the mirror controlled by this sensor and can lead to an instability. The coupling increases with the light power inside the interferometer. In GEO\,600 this coupling has been observed and would make the foreseen increase in laser power impossible. Operating the light source of the shadow sensor (an infra-red LED) with modulated current with frequencies in the kHz range and demodulating the signal of the photo-detector can make the system insensitive to low frequency light power fluctuations. If the system works well all local control systems of the main GEO optics will be converted to AC operation. First experiments have already shown a satisfactory noise performance.

\subsubsection{\label{TCS}Thermal Compensation System}
In contrast to the other gravitational wave detectors, which use resonators in the interferometer arms for power buildup, GEO\,600 achieves high circulating light power with a high Power Recycling factor of about 1000. As a consequence the full light power in the interferometer arms has to pass through the beam splitter. Although GEO\,600 uses fused silica glass with very low absorption of 0.5\,ppm/cm (Suprasil$^\registered$ 311SV), about 30\,mW are absorbed inside the beam splitter. The local temperature increase causes a thermal lens which limits the maximum power that can be used. A system shining infra-red light, which is absorbed by the beam splitter surface, will be used to reduce the thermal lensing. Instead of using a CO$_2$ laser like in the other gravitational wave detectors\cite{TCSLIGO},\cite{TCSVirgo} an incandescent light source will be used in GEO\,600 minimizing the fluctuations in heating power to the required level. In GEO\,600 the required power stability for the heating source is beyond the current possibilities of a CO$_2$ laser.

\section{Conclusions}
The upgrades for the GEO\,600 detector foresee a sensitivity enhancement for frequencies above 1\,kHz of about one order of magnitude within the next two years, i.e. until the middle of 2011.  In late January 2010 part of the upgrades have already been done: DC readout, Tuned Signal Recycling operation,and the Output Mode Cleaner are routinely operating. Locking stretches on the order of 10 hours have been shown. Squeezing is about to be installed in GEO\,600 in February 2010.\par
 As far as it can be foreseen now, in the period from 2011 to 2015 GEO600 will be the only interferometer in the world that can be on-line for astrophysical observations for a substantial amount of time. The three LIGO interferometers as well as Virgo will be upgrading to the advanced versions of their instruments. Therefore, current planning is, that GEO\,600 will spend more and more time in observation mode, as all the upgrades foreseen for GEO\,600 at this time are implemented, and the sensitivity would approach the planned goal.

\section{Acknowledgments}
The authors are grateful for support from the Science and Technology Facilities Council (STFC) in the UK, the BMBF, Max Planck Society (MPG) and the state of Lower Saxony in Germany. This work was partly supported by DFG grant SFB/Transregio 7 'Gravitational Wave Astronomy'. This document has been assigned LIGO Laboratory document number LIGO-P0900122.

\section*{References}

\begin{thebibliography}{9}
\bibitem{GEO600}H. Grote et al 2008 The status of GEO\,600, CQG 25 114043 
\bibitem{LIGO}R. Adhikari et al. 2009 LIGO Interferometers' Status \@ the Amaldi conference, submitted to CQG  
\bibitem{Virgo}B. Swinkels et al. 2009 Commissioning status of the Virgo interferometer \@ the Amaldi conference, submitted to CQG
\bibitem{HartmutGEO600Amaldi} H. Grote 2009 The GEO\,600 status, submitted to CQG
\bibitem{Titania}G. M. Harry 2007 Titania-doped tantala/silica coatings for gravitational-wave detection, CQG 24 405–415
\bibitem{LaguerreGauss}S. Chelkowski et al. 2009 Prospects of higher-order Laguerre-Gauss modes in future gravitational wave detectors, Phys. Rev. D 79, 122002
\bibitem{Nonquarterstacks}J. Agresti et al. Proc. 2006 Optimized multilayer dielectric mirror coatings for gravitational wave interferometers SPIE 6286 628608, LIGO-P060027-00-Z.
\bibitem{Schnupp} L. Schnupp 1988 Presentation at European Collaboration Meeting on Interferometric Detection of Gravitational Waves, (Sorrent, Italy, Oct 1988) (unpublished)
\bibitem{DCreadoutGEO} S.~Hild et al. 2009  DC-readout of a signal-recycled gravitational wave detector, CQG 26 055012
\bibitem{JeromeAmaldi} J. Degallaix 2009 Commissioning of the tuned DC readout at GEO600, JPCS, published in this issue
\bibitem{OMCMirko} M. Prijatel et al. Control and automatic alignment of the output mode
cleaner of GEO600, published in this issue
\bibitem{coreoptics}W. Winkler et al. 2007, The core optics of GEO600, Optics Communications 280, 491-499
\bibitem{Buonanno}A. Buonanno et al 2003 Quantum noise in laser-interferometer gravitational-wave detectors with a heterodyne readout scheme Phys. Rev. D 67 122005
\bibitem{SqueezingChel}S. Chelkowski et al. 2005 Experimental characterization of frequency-dependent squeezed light, Phys. Rev. A {\bf 71} 013806
\bibitem{FilteredSqueezing}J. Harms et al. 2003 Squeezed-input, optical-spring, signal-recycled gravitational-wave detectors, Phys. Rev. D {\bf 68}, 042001
\bibitem{TwinSigRec}A.~Th\"uring, et. al, broadband queezing of quantum noise in a Michelson interferometer with Twin-Signal Recycling, Optics Letters, 34, (6), (2009)
\bibitem{TunedSR}S. Hild et al. 2007 Demonstration and comparison of tuned and detuned signal recycling in a large-scale gravitational wave
detector, CQG 24 1513
\bibitem{EnhLIGO}J. Smith 2009 The path to the enhanced and advanced LIGO gravitational-wave detectors, CQG 26 114013
\bibitem{Virgo+}F Acernese et al. 2008 The Virgo 3 km interferometer for gravitational wave detection, J. Opt. A: Pure Appl. Opt. 10 064009
\bibitem{GEOsusp} B. Willke et al. 2002 The GEO\,600 gravitational wave detector, CQG 19 (7) 1377-1387
\bibitem{TCSLIGO} S. J. Waldman, 2006 Status of the LIGO at the start of the fifth science run, CQG 23 S653–S660
\bibitem{TCSVirgo} F. Acernese et al. 2008 Status of Virgo, CQG 25 114045
\end{thebibliography}

\end{document}